**(Extended Abstract)**
# Mathematical Optimization of Resolution Improvement in Structured Light data by Periodic Scanning Motion: Application for Feedback during Lunar Landing

**Tarek A. Elsharhawy[1,2], P. James Schuck[1], Shuo Liu[3] and Luc Saikali[2]**

This Article / research proposes the development and mathematical optimization of a sophisticated Structured Light System (SLS) integrated with Structured Illumination Microscopy (SIM) techniques (moiré pattern, see [1-2]) for providing real-time feedback during lunar landings. The research focuses on transcending the current limitations of resolution enhancement in conventional SLS by employing advanced SIM principles, combined with advanced Iterative Learning Control (ILC) methodologies (see [3-12]). The proposed system aims to provide a significant enhancement to current obstacle avoidance systems, such as LiDARs, which are commonly used in lunar landing scenarios but may face limitations in resolution and precision under low-light and high-noise conditions. By integrating SIM with SLS, this research seeks to offer a more robust solution that can work in tandem with LiDAR systems, providing the necessary resolution improvements for precise lunar landings.

The preliminary phase of this research involved the development of MATLAB model, named SIMworldmodel17, which serves as the foundational step toward achieving the objectives of the proposed system. This MATLAB script is specifically designed to optimize parameters related to SIM technique (moiré pattern), enabling users to experiment with and refine the parameters that influence image resolution. The code develops a graphical user interface (GUI) that allows users to interactively adjust key parameters such as the number of angles used in the structured light, the Gaussian sigma value for smoothing, and the neighborhood size for Wiener filtering. Additionally, the GUI includes a checkbox for toggling the application of smoothing, providing a versatile platform for conducting optimization experiments.

The structure of SIMworldmodel17 (Figure 1) is centered around the interactive GUI, which is divided into multiple subplots for displaying both the original and restructured high-resolution images, as well as additional plots for further analysis. Users can manipulate the parameters through sliders and checkboxes, with real-time updates reflected in the displayed images and plots. This feature is particularly beneficial for researchers as it allows them to immediately visualize the effects of their adjustments, fostering a more intuitive understanding of the parameter space and the optimization process.

[1] Doctoral Candidate, Columbia University, School of Engineering & Applied Science, Columbia University, 500 West 120th Street, New York  NY 10027, USA.

[2] Lecturer, Cal Poly Pomona, College of Engineering, 3801 W. Temple Ave., Pomona, CA 91768 USA.

[1] Professor / Doctoral Advisor, Mechanical Engineering Department. Columbia University, 500 West 120th Street, New York  NY 10027, USA.

[3] Doctoral Candidate, Department of Mechanical Engineering, Boston University, 110 Cummington Mall, Boston, MA 02215.

[2] Graduate Student, Department of Computer Science, Cal Poly Pomona, 3801 W. Temple Ave., Pomona, CA 91768 USA.

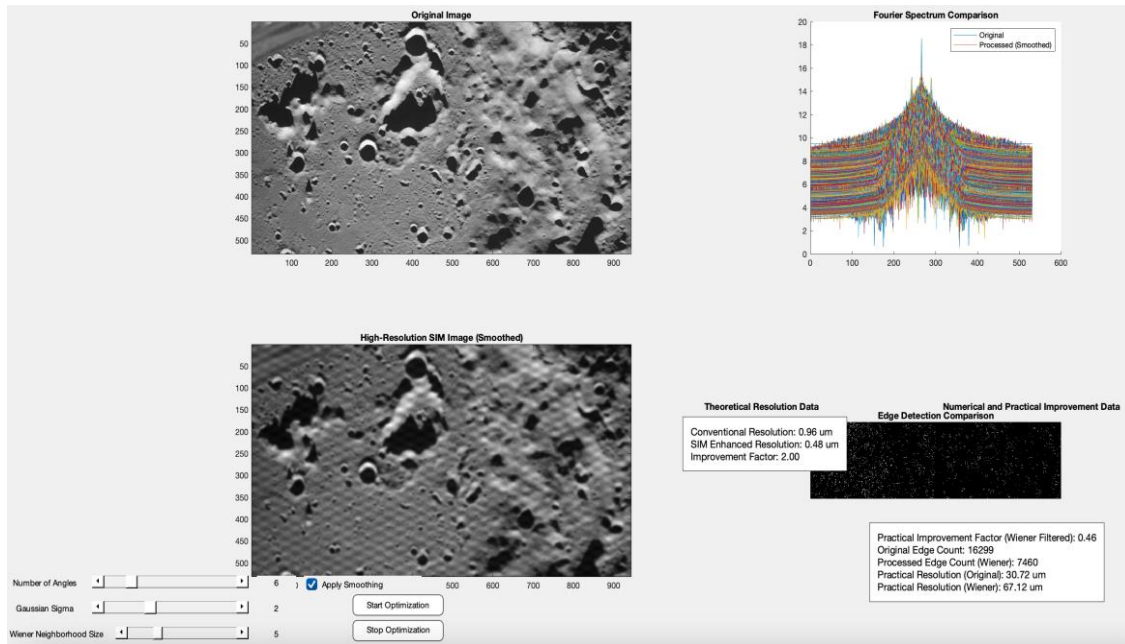

**Fig. 1** SIMworldmodel17

Once the user has configured the desired parameters, the GUI includes a button to start the optimization process. This process involves running a series of algorithms designed to enhance image resolution based on the input parameters. The MATLAB code, therefore, not only facilitates the manual adjustment of parameters but also automates the optimization process, providing users with a powerful tool for exploring the full potential of SIM techniques in SLS configuration.

Initial results obtained using SIMworldmodel17 have been promising, demonstrating the capability of this approach to significantly enhance image quality by optimizing the structured light parameters. The interactive nature of the GUI allows for iterative refinement, enabling users to explore a wide range of parameter combinations to identify the optimal settings for their specific application. The results indicate that by fine-tuning the number of angles, adjusting the Gaussian sigma for smoothing, and carefully selecting the Wiener filtering neighborhood size, it is possible to achieve substantial improvements in image resolution. This capability is particularly critical for applications such as lunar landings, where detailed terrain mapping is essential for safe and precise navigation.

In the context of lunar landings, the integration of SIM techniques with traditional SLS systems represents a novel approach to addressing the challenges posed by low-light and high-noise environments. Lunar landings, especially during low light, present significant difficulties due to the lack of natural light and the presence of uneven terrain.

Traditional SLS, while effective in certain conditions, often fall short in these challenging environments. By incorporating SIM, this research aims to overcome these limitations, providing a means to enhance the resolution and clarity of images captured during lunar descent. Moreover, this enhanced SLS could act as a complementary system to LiDARs, filling in the gaps where LiDARs struggle with resolution and precision, particularly in environments where clear visual feedback is critical.

One of the key innovations of this research is the application of ILC within the SIMworldmodel17 framework. ILC is a control strategy designed to improve system hardware performance through repeated iterations, learning from previous errors to refine the output in subsequent trials. In the context of SIMworldmodel17, ILC is used to optimize the periodic scanning motions of the SLS, ensuring that the system hardware operates at its peak performance even under the harsh conditions of space. By iteratively adjusting the scanning motions based on feedback from the system, ILC can help to minimize errors and maximize the resolution enhancement achieved by the SIM techniques.

The effectiveness of the proposed system will be validated through a series of simulations and experiments. These simulations will be designed to mimic the conditions of lunar descent, including variations in terrain, height, lighting, and other environmental factors. The goal is to assess the performance of the system under these conditions, determining its ability to provide high-resolution, real-time feedback that is critical for safe and precise lunar landings. The experiments will focus on evaluating the resolution enhancement achieved by the system, as well as its ability to adapt to changes in the environment through the use of ILC. The MATLAB code developed in this research, SIMworldmodel17, represents the first step towards a more sophisticated world model, laying the groundwork for future advancements in real-time terrain mapping and navigation systems for space exploration.

The integration of SIM techniques, SLS, and ILC represents a significant advancement in the field of lunar landing technology. The research aims not only to improve the resolution of images captured during lunar descent but also to enhance the overall reliability and safety of lunar landings. By providing real-time feedback and high-resolution terrain mapping, the proposed system could play a crucial role in future lunar missions, particularly those that involve landing in scientifically valuable but challenging terrains. Furthermore, its potential to work in conjunction with existing LiDAR systems could create a more comprehensive obstacle avoidance and navigation system, reducing the risks associated with lunar landings.

In addition to its applications in lunar landings, the technology developed in this research has potential implications for other real-time navigational applications that require enhanced visual feedback systems in challenging environments. For example, the system could be adapted for use in other planetary exploration missions, where the ability to navigate accurately in low-light conditions is essential. This system is designed to potentially have the capability to land on terrestrial surfaces without prior knowledge of the terrain and without the need for extensive training data. It could also be applied in terrestrial applications, such as autonomous vehicles or drones operating in complex or poorly lit environments."

In conclusion, this research lays the groundwork for a revolutionary approach to lunar landing navigation by combining SIM techniques with SLS and ILC. The preliminary MATLAB code, SIMworldmodel17, and its results demonstrate the feasibility and potential of this approach. This MATLAB code is the first step towards a more sophisticated world model, designed to optimize SIM parameters in real-time, coupled with the iterative learning capabilities provided by ILC. This combination offers a powerful tool for enhancing image resolution in SLS. This research is expected to contribute significantly to the advancement of lunar landing technologies, providing a robust and reliable system for real-time feedback and high-resolution terrain mapping during lunar descent. Furthermore, the system is designed to have the capability to land on terrestrial surfaces without prior knowledge of the terrain and without the need for extensive training data. The ultimate aim is to deliver a system that can significantly improve the accuracy and reliability of lunar landings, contributing to the success of future space missions and potentially influencing a wide range of other real-time navigational applications. The proposed system's synergy with existing technologies like LiDAR further underscores its importance, promising to enhance overall obstacle avoidance capabilities and ensure safer, more accurate landings in challenging extraterrestrial environments.